# Quantifying Ionic Liquid Affinity and Its Effect on Phospholipid Membrane Structure and Dynamics

V. K. Sharma[1,2*], J. Gupta[1,2], H. Srinivasan[1,2], P. Hitaishi[3$], S. K. Ghosh[3], S. Mitra[1,2]

[1]*Solid State Physics Division, Bhabha Atomic Research Centre, Mumbai, 400085, India*

[2]*Homi Bhabha National Institute, Mumbai, 400094, India*

[3]*Department of Physics, School of Natural Sciences, Shiv Nadar Institution of Eminence, NH 91, Tehsil Dadri, G. B. Nagar, Uttar Pradesh  201314, India*
[$]*Institute for Experimental and Applied Physics, Christian-Albrechts-University of  Kiel, 24118 Kiel, Germany (present address)*

## Abstract

Understanding the interactions between ionic liquids (ILs) and biomembranes is pivotal for uncovering the origins of IL-induced biological activities and their potential applications in pharmaceuticals. In this study, we investigate the influence of imidazolium-based ILs on the viscoelasticity, dynamics, and phase behavior of two model membrane systems: (i) lipid monolayers and (ii) unilamellar vesicles, both composed of dipalmitoylphosphatidylcholine (DPPC). Two different ILs with varying alkyl chain lengths, namely, 1-decyl-3-methylimidazolium bromide (DMIM[Br]) and 1-hexyl-3-methylimidazolium bromide (HMIM[Br]) are used to investigate the role of alkyl chain lengths. Our findings demonstrate that both ILs induce significant disorder in lipid membranes by altering the area per lipid molecule, thereby modulating their viscoelastic properties. IL with longer alkyl chain shows stronger interactions with membranes, causing more pronounced disorder. Fourier-transform infrared spectroscopy indicates that IL incorporation shifts the membrane's main phase transition to lower temperatures and introduces gauche defects, signifying increased structural disorder. This effect is amplified with longer alkyl chains and higher IL concentrations. Quasielastic neutron scattering studies highlight that ILs markedly enhance the lateral diffusion of lipids within the membrane leaflet, with the extent of enhancement determined by the membrane's physical state, IL concentration, and alkyl chain length. The most pronounced acceleration in lateral diffusion occurs in ordered membrane phase with higher concentrations of the longer-chain IL. Molecular dynamics simulations corroborate these experimental findings, showing that longer-chain ILs extensively disrupt lipid organization, introduce more gauche defects, increase the area per lipid, and consequently enhance lateral diffusion. This increase in lipid fluidity and permeability provides a mechanistic basis for the observed higher toxicity associated with longer-chain ILs. These results offer critical insights into the molecular-level interactions of ILs with lipid membranes, advancing our understanding of their toxicological and pharmaceutical implications.

*Corresponding Author: Email: sharmavk@barc.gov.in;  Phone +91-22-25594604

# 1. INTRODUCTION

Ionic liquids (ILs) are compounds entirely composed of ions, characterized by melting points below 100°C[1]. ILs possess exceptional physicochemical properties and biological activities, making them widely applicable across diverse industries, including catalysis, electrochemistry, and pharmaceuticals[2-3]. They are extensively utilized for drug solubilization, protein and nucleic acid stabilization, and the development of advanced biosensors. However, their water solubility and limited biodegradability have raised significant environmental concerns, particularly due to their potential toxicity to living organisms[4-6]. Intriguingly, ILs have demonstrated potent antimicrobial properties, effectively inhibiting bacterial growth[7-8]. These biological activities are believed to arise from ILs' strong interactions with cell membranes[5, 8-12]. However, the precise molecular mechanisms underlying these interactions remain a subject of active research. In addition to the antimicrobial and cytotoxic effects of ILs on membranes, understanding their interactions with lipid membranes is crucial for optimizing and fine-tuning drug delivery systems[13]. Liposomes, extensively studied over the past few decades, have emerged as promising carriers for drug and gene delivery[13]. With structural and compositional similarities to cell membranes, liposomes can fuse with them, enabling the targeted release of their contents into the cytoplasm. These nanoparticles have played a transformative role as drug nanocarriers, exemplified by their success in mRNA-based COVID-19 vaccines[14], highlighting their pivotal role in the ongoing bio-nanotechnology revolution. However, a key challenge remains: achieving precise control over drug release from vesicles. Addressing this challenge requires a deeper understanding of the factors influencing lipid membrane properties[15].

Cell membranes exhibit remarkable viscoelastic properties, balancing rigidity for structural integrity with flexibility for dynamic processes like endocytosis and exocytosis. At their core, lipid bilayers are highly fluid and dynamic, with continuous fluctuations in lipid positions, orientations, curvature, and thickness[7, 16-19]. The dynamics and viscoelasticity of membrane play a pivotal role in physiological functions such as cell signaling, membrane-protein interactions, cellular shape modulation, fusion and fission, and selective permeability. Cell membranes must maintain balanced viscoelasticity and fluidity to perform their essential functions effectively. The fluidity of a membrane describes the degree of mobility and flexibility exhibited by its constituents within the bilayer. Viscoelastic properties of membranes enable them to endure mechanical stresses and deformations while preserving their structural integrity and shape. Similarly, membrane fluidity, a fundamental property dictated by the dynamic behavior of lipids within the bilayer, is intricately linked to critical membrane processes. Any disruption in these properties can compromise membrane functionality, ultimately affecting cellular stability and viability. Thus, a comprehensive understanding of how ILs impact the viscoelasticity, phase behavior, and dynamics of cell membrane is essential.

The cell membrane is a complex, heterogeneous assembly of lipids, proteins, carbohydrates, and small molecules. Lipid bilayers, as simplified model systems, provide an ideal framework for probing IL-membrane interactions. The toxicity of these ILs strongly depends on their alkyl chain lengths[5]. It has been suggested that penetration of IL in lipid membrane is driven by favourable hydrophobic interactions between the alkyl chains of lipids and ILs[20]. Strength of hydrophobic interaction can be easily tuned by changing the chain length of IL, hence the alkyl chain length of ILs is a crucial parameter to consider[21-22]. Another important parameter is the concentration of ILs. It has been shown that the permeability of membrane increases with rising the concentration and chain length of IL[23]. Increased permeability is directly associated with cytotoxicity, as it signifies the leakage of cellular contents or may ultimately lead to complete disruption. Hence, alkyl chain length and concentration of ILs are two important parameters to study their influence on the biophysical properties of the membrane.

Surface area-pressure (A - π) isotherm and rheology measurements are widely used to investigate lipid monolayers[24] . Isotherm measurements provide information about the area per molecule, occupied by the lipid molecules in the monolayer and the thermodynamics of a film. Interfacial rheology measurements elucidate the viscoelastic properties of the monolayer. Zeta potential[22] measurements is a highly effective method in assessing the binding affinity of ILs with lipid membranes. Differential scanning calorimetry (DSC) has been widely employed to examine the influence of ILs on the phase behavior of lipid membranes[23], while Fourier-transform infrared (FTIR) spectroscopy provides crucial insights into phase transitions and conformational changes in lipid molecules[18]. These methods are highly complementary to each other and provide information about the enthalpy, and conformational changes occurring in the lipid molecules, during the phase transitions. Quasielastic neutron scattering (QENS) is a powerful technique for probing lipid dynamics across time scales ranging from subpicoseconds to nanoseconds and spatial scales from angstroms to nanometers [25-32]. As a scattering-based method, QENS uniquely provides simultaneous insights into both the temporal and spatial aspects of molecular motions. This capability makes it an indispensable tool for deciphering the intricate, multi-scale dynamics of lipid membranes

In this study, dipalmitoylphosphatidylcholine (DPPC), a saturated phosphatidylcholine lipid with a well-defined main phase transition temperature ~ 41°C, was selected as the model lipid membrane system. The main phase transition temperature of DPPC is experimentally convenient, allowing investigations into the influence of the membrane's physical state on its interactions with ILs. Imidazolium-based ILs were chosen for their exceptional physicochemical properties, ease of synthesis, and remarkable biological activity, making them among the most extensively studied IL families. To examine the role of alkyl chain length on IL-membrane interactions, two representative ILs were selected: 1-decyl-3-methylimidazolium bromide (DMIM[Br] or $C_{10}MIM[Br]$) and 1-hexyl-3-methylimidazolium bromide (HMIM[Br] or $C_6MIM[Br]$).

A comprehensive understanding of how ILs modulate the viscoelasticity, phase, and dynamic behavior of lipid membranes can offer valuable insights into their broader biological implications. This study employs a suite of biophysical techniques such as Langmuir trough for pressure-area isotherms and dilation rheology to assess membrane mechanics, zeta potential to quantify binding affinity, FTIR spectroscopy to explore phase transitions and conformation change in the lipids, and QENS to investigate membrane dynamics at nanoscales. To get microscopic insights, we have also carried out molecular dynamics (MD) simulation. These complementary approaches enable a holistic view of how ILs impact membrane integrity, fluidity, and functionality, paving the way for advancements in both the safe use of ILs and their optimization as tools for targeted drug delivery.

## 2. MATERIALS AND METHODS

### 2.1 Materials

The phospholipid DPPC was purchased in powder form from Avanti Polar Lipids (Alabaster, AL). The ionic liquids, HMIM[Br], DMIM[Br] along with $D_2O$ were obtained from TCI Chemical or Sigma Aldrich. HPLC spectroscopy grade chloroform (purity > 99.9%) was purchased from Sigma Aldrich.

## 2.2 Sample Preparations

Two distinct model systems namely, monolayers and unilamellar vesicles (ULVs), were employed to explore the interactions between ILs and lipid membranes. For monolayer studies, stock solutions of DPPC, HMIM[Br], and DMIM[Br] were prepared individually in chloroform at a concentration of 0.5 mg/mL. To achieve the desired weight percentage (wt%) of ILs, precise volumes of IL and DPPC solutions were mixed. This mixture was then used to prepare lipid monolayers, which were systematically characterized through pressure-area isotherm measurements and in-plane viscoelasticity studies. For ULVs preparation, the extrusion method was employed as follows: DPPC lipid powder was dissolved in chloroform, which was then evaporated under a gentle stream of nitrogen gas. To remove any residual organic solvent, the vial was placed under vacuum ($10^{-3}$ atm) overnight at 330 K. The resulting dry lipid films were hydrated at 330 K and vortex-mixed for 10 minutes. Vesicles were then extruded by passing the suspension through a mini-extruder (Avanti Polar Lipids, Alabaster, AL) equipped with a polycarbonate membrane (average pore diameter of 100 nm) more than 31 times. The extruder was placed on a hot plate set at 330 K to ensure the DPPC membranes remained in the fluid phase during extrusion. To incorporate HMIM[Br] and DMIM[Br] ILs into the lipid membranes, the appropriate amounts of ILs were added to the vesicle solution at 330 K and thoroughly mixed. The mixture was allowed to equilibrate at 330 K for approximately 30 mins before being used for measurements.

## 2.3 Surface Area-Pressure Isotherm

Thermodynamic interactions of an IL with a DPPC lipid membrane have been quantified using surface area-pressure isotherms, in-plane static elasticity and dilation rheology. A Langmuir-Blodgett (LB) trough of size $36.4 \times 7.5 \times 0.4$ cm$^3$ (KSV NIMA, Biolin Scientific) with two symmetric Delrin barriers and a platinum Wilhelmy balance was used to record the isotherms of a monolayer. The lipid and ILs were individually dissolved in chloroform to prepare a stock solution with a concentration of 0.5 mg/ml. Later the solutions of IL and lipid were combined to achieve ~ 4 and 10 wt % ILs. The temperature of subphase filled in the trough was maintained to 300 K by circulating water through a hollow panel below the trough using a water bath (Equibath, India). This temperature is below the main phase transition temperature ($T_m$) of the DPPC, therefore, the lipids in the monolayer at air-water interface remain in the gel phase. To generate an isotherm, the surface pressure was monitored as a function of the mean molecular area. A specific volume of lipid or lipid/IL solution was spread on the water surface of the trough using a glass Hamilton micro syringe, followed by 20-minute waiting time to allow the complete evaporation of chloroform. The monolayer was compressed at a consistent rate of 4 mm/min until it reached the collapse pressure.

The in-plane elasticity, also referred to as compressional modulus, of the monolayer is calculated from the area-pressure isotherms data using the equation[33],

$$E = -A\left(\frac{\partial \pi}{\partial A}\right)_T \tag{1}$$

Here, $A$ denotes the mean molecular area whereas $\pi$ is the lateral surface pressure at a given temperature ($T$). This is known as static elasticity as the slow compression of monolayer ensures the process to be a quasistatic one.

### 2.4 Dilation Rheology of Lipid Monolayer

In-plane rheology measurements were conducted using a Langmuir-Blodgett trough, where a monolayer was slowly compressed at a rate of 2 mm/min to a target surface pressure. Subsequently, the two symmetrical barriers were oscillated at a specific frequency. The resulting time-dependent strain in the area was recorded, along with the corresponding stress response in the form of time-dependent surface pressure. The strain magnitude applied was 1% relative to the surface area of the monolayer film, and the oscillation frequency of the barriers ranged from 70 to 370 mHz.

The following equations were used to estimate the dynamic elasticity of the monolayer[34-35],

$$a(t) = A(1 + a_0 \sin(\omega t + \phi_a) \tag{2}$$

$$\pi(t) = P + \pi_0 \sin(\omega t + \phi_\pi) \tag{3}$$

where $A$, $a_0$, and $a(t)$ are the initial surface-area, amplitude of applied strain (1% of $A$) and time dependent change in the area, respectively. $\omega$ is the angular frequency of oscillation, $\pi_0$ is the stress amplitude and $P$ is the initial surface pressure. $\pi(t)$ is the time-dependent response in monolayer surface pressure. Because of the viscoelastic nature of the film, there is a phase lag between stress and strain, which is given by $\phi = \phi_\pi - \phi_a$. The two viscoelastic parameters of the monolayer, namely, the storage modulus ($E^{'}$) and loss modulus ($E^{''}$) are expressed in term of this measured phase lag by the equations,

$$E^{'} = \frac{\pi_0}{a_0} \cos \phi$$
$$E^{''} = \frac{\pi_0}{a_0} \sin\phi \tag{4}$$

While the storage modulus ($E^{'}$) is related to the elastic nature of the monolayer, the loss modulus ($E^{''}$) is connected to the viscous nature of the film.

## 2.5 ζ-Potential

To assess the binding affinity of ILs with the lipid membrane, ζ-potential measurements were conducted on 0.2 mg/ml DPPC ULVs in PBS at varying concentrations of HMIM[Br] and DMIM[Br] at 320 K. Each sample was thermally equilibrated for 5 minutes before measurement. The ζ-potential measurements were performed using the Zetasizer Nano ZS system (Malvern Instruments, U.K.), which was equipped with a 633 nm He–Ne laser.

## 2.6 FTIR

Infrared spectroscopic studies were conducted on 5 wt % DPPC ULVs in $D_2O$ under varying concentrations of HMIM[Br] and DMIM[Br] at different temperatures. The measurements were carried out using a Bruker IFS125 Fourier Transform spectrometer, equipped with a liquid nitrogen-cooled MCT detector, a KBr beam splitter, a globar source, and a temperature-controlled sample stage. The liquid cell sample chamber consisted of two $CaF_2$ windows (10 mm diameter) with an approximate path length of ~60 µm. For each measurement, 40 µL of the ULV sample was carefully loaded into the chamber, which was continuously purged with dry nitrogen gas to eliminate moisture interference. All experiments were conducted in transmission mode, with spectral data converted to absorbance using the relation $Ab = -ln(Tr)$ where $Ab$ represents absorbance and $Tr$ is the transmittance. Data acquisition covered the mid-infrared range (900–4000 cm$^{-1}$) across a temperature span of 303–328 K, with increments of 1 K/min. Background correction involved the subtraction of spectra from an empty cell and $D_2O$. Each spectrum, collected with a minimum resolution of 2 cm$^{-1}$, was averaged over 120 scans for enhanced signal-to-noise ratio. Spectral analysis was performed using Lorentzian line-shape fitting to extract precise vibrational features of the system.

## 2.7 QENS

Quasielastic neutron scattering (QENS) experiments were carried out using a high energy resolution spectrometer IRIS, at the ISIS pulsed Neutron and Muon source at the Rutherford Appleton Laboratory, UK. IRIS is an inverted geometry backscattering spectrometer, with a PG(002) analyzer, offers an energy resolution ΔE = 17 μeV (full width at half-maximum). IRIS was used in the offset mode with an accessible range of energy transfer from -0.3 to +1.0 meV and the available $Q$-range from 0.5 to 1.8 Å$^{-1}$. QENS experiments were performed on 5 wt % DPPC vesicles with varying concentrations of DMIM[Br], namely 0, 10 wt % and 20 wt % with respect to DPPC. To investigate the effects of chain length of IL, QENS measurements were also carried out on DPPC vesicles with 10 wt % HMIM[Br]. Deuterated water was used as a solvent to minimize the scattering contribution other than lipids, because the neutron scattering cross-section of hydrogen is an order of magnitude higher than deuterium. To investigate the role of physical state of the membrane, QENS measurements were carried out at 310 K (below the $T_m$ of DPPC) and 345 K (above the $T_m$ of DPPC). Samples were placed in annular aluminium sample cans with 0.5 mm internal spacing such that the sample scattering was no more than 10%, thereby minimizing multiple scattering effects. Instrument resolution was obtained by measuring the QENS data from a standard vanadium sample. To estimate the solvent contribution, QENS data were recorded on pure $D_2O$ at 310 K and 345 K. MANTID Software[36] was used for standard data reduction including background subtractions.

## 2.8 MD Simulation

The initial configuration of the DPPC lipid bilayer, comprising of 128 lipids (64 per leaflet), was constructed using CHARMM-GUI[37]. The system was solvated with a water-to-lipid ratio of 111:1. The structures of DMIM$^+$ ions were generated and optimized using Avogadro software[38]. To simulate interactions, DMIM$^+$ and Br$^-$ ions were placed at least 5 Å from both the upper and lower leaflets of the equilibrated bilayer, with an equal distribution above and below the bilayer. The CHARMM36 force field[39] and TIP3P[40] water model were employed, with DMIM$^+$ parameters generated via the CGenff server[41]. All-atom MD simulations, which account for all pairwise atom-atom interactions, were performed in the NPT ensemble using a Langevin barostat and thermostat. Long-range interactions were treated with the particle mesh Ewald method, employing a real-space cutoff of 12 Å, and the integration time step was set at 1 fs. MD simulations were carried out at three different concentrations of DMIM[Br] namely 0, 10, and 25 mol % which are equivalent to 0, 4, and 10 wt %, respectively. MD simulations were carried out at 7 different temperatures namely 326 K, 322 K, 318 K, 314 K, 310 K, 307 K, 303 K in the cooling cycles. Similar to experiments, first system equilibrated in the fluid phase for about 100 ns which allows most of ILs to be inserted in the lipid membrane then it is cooled through simulated annealing at a rate of -1 K/ps. After cooling them down to 303 K, the system was further equilibrated for 100 ns in the ordered phase. The final production runs, lasting 100 ns with atom positions recorded every 10 ps, were analyzed to gain insights into the system's dynamics. All simulations were conducted using the NAMD simulation package[42].

## 3. RESULTS AND DISCUSSION

### 3.1 Stronger Binding Affinity of Longer Chain IL with Lipid Membrane

To investigate the binding behavior of ILs, the $\zeta$-potentials of DPPC ULVs were measured at varying HMIM[Br] and DMIM[Br] concentrations. Figure 1 illustrates the changes in the $\zeta$-potential of DPPC vesicles upon addition of increasing concentrations of both ILs at 320 K. In the fluid phase at pH ~7.4, zwitterionic DPPC vesicles in PBS buffer exhibit a negative $\zeta$-potential [43]. The binding of IL cations to the vesicle surface results in a shift of the $\zeta$-potential towards less negative or even positive values. The extent of this shift provides valuable insight into the IL affinity to membrane and loading capacity of the lipid bilayers for a given hydrophobic cation. As the IL cations partition into the membrane, the negative $\zeta$-potential of the DPPC vesicles decreases. For HMIM[Br], the $\zeta$-potential shows only a modest increase, remaining negative even at concentrations as high as 11 mM. In contrast, the longer-chain IL, DMIM[Br], causes a substantial rise in the $\zeta$-potential, with a complete reversal to positive values observed at concentrations as low as 0.2 mM. Starting from an initial $\zeta$-potential of -10.5 ± 1.6 mV, the $\zeta$-potential of ULVs increased by $\Delta\zeta$ = 29.3 ± 1.8 mV upon the addition of 2 mM DMIM[Br], indicating strong binding between DMIM$^+$ and the DPPC membrane. In contrast, the addition of HMIM[Br], even at higher concentrations (up to 11 mM), resulted in only a slight increase of 2.6 ± 2 mV in the ζ-potential. This suggests that ILs with longer alkyl chains, such as DMIM[Br], exhibit a significantly stronger binding affinity for lipid membranes. These results underscore the critical role of hydrophobic interactions in IL-lipid membrane binding and highlight how alkyl chain length modulates the strength of these interactions. Results are consistent with isothermal titration calorimetry (ITC) studies[7] suggesting that the primary driving force for the incorporation of amphiphilic ILs into zwitterionic lipid bilayer is the hydrophobic effect. This hydrophobic effect arises from the unfavorable entropy of water molecules interacting with the hydrophobic regions of the ILs. In case of HMIM[Br], short alkyl chain fails to generate sufficient hydrophobic binding free energy to overcome the entropic penalty associated with transferring the amphiphile from the aqueous phase to the membrane interface.

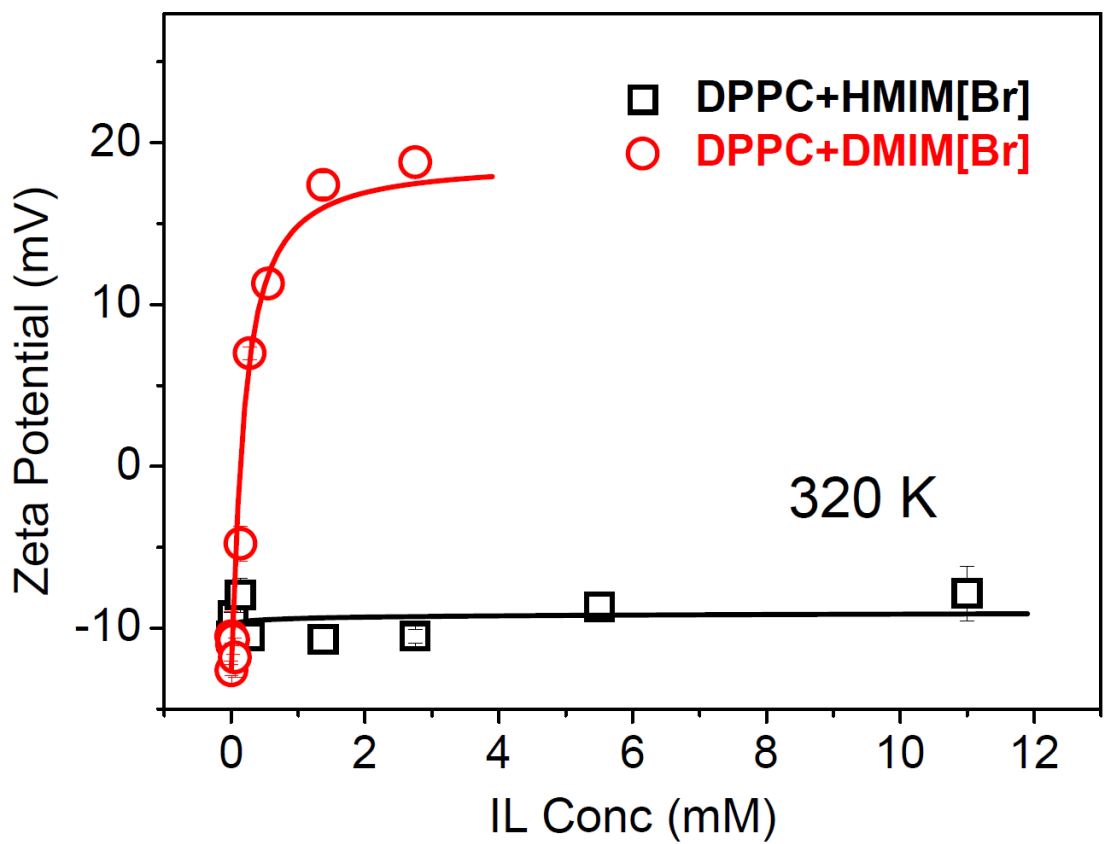

**FIGURE 1**: ξ-potentials of DPPC ULVs as a function of IL concentrations. Solid lines represent the best fit to the experimental data using Eq. (5).

Based on ζ-potential measurements, the binding constants of ILs could be estimated using a simple model that consider changes in the surface potential due to binding of the charged amphiphiles. In this model, ζ-potential can be written as[44]

$$\xi = \frac{\xi_o + \xi_{max}(KC)^n}{1 + (KC)^n} \qquad (5)$$

where, $\xi_o$ is the zeta potential of liposomes in the absence of ILs, $\xi_{max}$ is the zeta potential value for liposomes saturated with ILs, C is the concentration of IL, $K$ is the association constant and $n$ is stoichiometry (no of IL/no of lipids). Eq. (5) is used to describe the ζ-potential, where $K$ and $n$ are fitting parameters. For DMIM[Br], association constant $K$ and stoichiometry ($n$) are found to be $5.3\pm0.4\times10^3$ M$^{-1}$ and $1.2\pm0.1$. For shorter chain length IL, HMIM[Br], $K$ and $n$ are found to be $0.1\times10^3$ M$^{-1}$ and $0.1$. This suggests that longer chain IL has a stronger binding affinity with lipid membrane which is consistent with ITC results[7, 21].

One can estimate Gibbs free energy associated with the binding of IL from the association constant (K) using $\Delta G = -RT \ln 55.5K$ and are found to be -8.0 Kcal/mol and -5.5 Kcal/mol for DMIM[Br] and HMIM[Br], respectively. Present measurements suggest that as the hydrophobic chain lengths of IL increase it leads to stronger binding with the lipid membrane. To elucidate the role of the binding affinity of ILs on the structural, dynamical, and phase behavior of membranes, we conducted pressure-area isotherm, dilation rheology, FTIR, and QENS measurements. The findings are presented and discussed in the following sections

## 3.2 Ionic Liquid Enhances the Mean Lipid Area and Reduces the Elasticity of Molecular Film

In the present study, the monolayer is formed by homogenous mixture of DPPC and IL which was spread on the water subphase. Isotherms are generated by recording the variation in surface pressure on compression of the layer by bringing the barriers close to each other. The surface area-pressure isotherms of DPPC lipid with added ILs are shown in Fig. 2. Here, in all cases, the area per molecules (APM) is calculated by considering only the number of DPPC molecules. Therefore, the shift of the isotherm to a higher APM is the signature of the interaction of the ILs with lipid monolayer. Note that the concentration of ILs in DPPC/IL mixed system is increased in such a way that the number of lipid molecules spread on the interface remains constant.

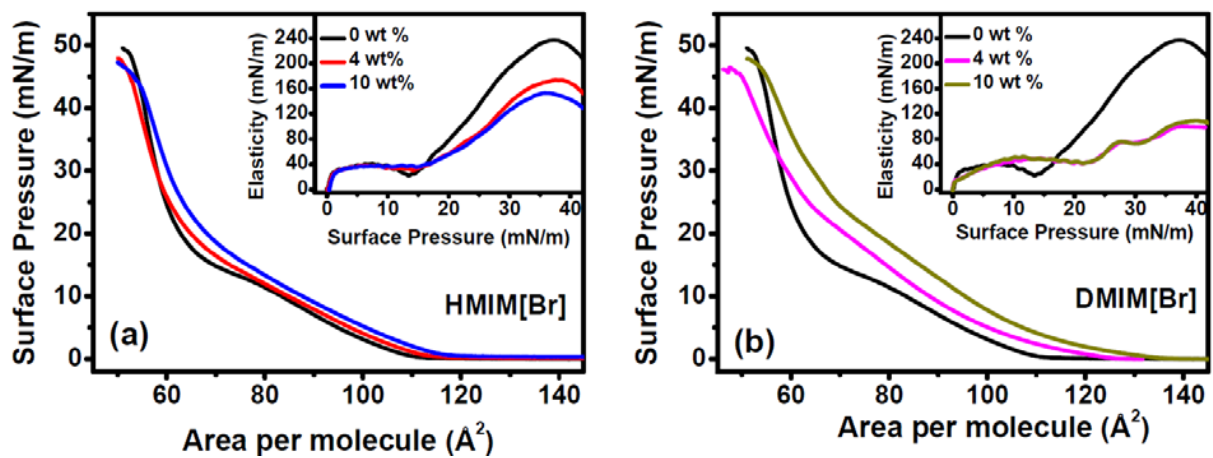

**FIGURE 2:** Surface area-pressure isotherms of monolayers deposited at air-water interface, composed of DPPC lipid with added 0, 4 and 10 wt% ILs, (a) HMIM[Br] and (b) DMIM[Br]. All measurements are performed at 300 K and inset figures exhibits the corresponding in-plane elasticity of respective monolayers.

Initially, the barriers are at an extreme position with the molecules being far apart from each other. Due to negligible intermolecular interaction, there is no measurable surface pressure. This region is referred to as gaseous (G) phase. Compression of barriers brings the molecules close enough so that they start interacting, showing an increase in surface pressure. This region is known as liquid extended (LE) phase. The maximum area at which the surface pressure shows the first non-zero value is known as lift-off area ($A_L$). It is evident that the $A_L$ of DPPC is increased slightly from 110 to 113 and to ~117 Å$^2$ for 4 and 10 wt% of the HMIM[Br] in DPPC monolayer, as shown in Fig. 2 (a). This increase in $A_L$ is much more prominent in the presence of DMIM[Br] with the values of 124 and 132 Å$^2$ at the same concentration of the IL, as shown in Fig. 2 (b). This is quite conclusive that the DMIM[Br] affects the lipid monolayer to a greater extent. On further compression, the monolayer undergoes a transformation from LE to liquid condensed (LC) phase, where molecules are in a compact arrangement. For the pure DPPC lipid, there is an intermediate plateau region between the LE and LC phases suggesting the coexisting of these phases. The plateau region shifts toward the higher surface pressure in the presence of both the ILs. This is clearly seen as a dip in the in-plane elasticity curve, shown in the inset of

Fig. 2. The shift of the dip to the higher surface pressure in presence of the IL is more prominent for DMIM[Br]. Extension of the LE phase up to a higher surface pressure is an indication of the disordering effect of the IL in the lipid monolayer. This is further manifested in the drop of the elasticity of the lipid layer. The long chain IL reduces the elasticity more and made the membrane more flexible. Such an observation of shifting the isotherm towards higher area with a higher interaction with increased alkyl chain length of IL is reported earlier[45]. The area occupied by a molecule is a critical factor that influences the thickness, flexibility, and permeability of a lipid membrane. The impact of IL on this lateral area of the molecule suggests that it could disrupt the physical properties of a cellular membrane, potentially affecting its functionality[46].

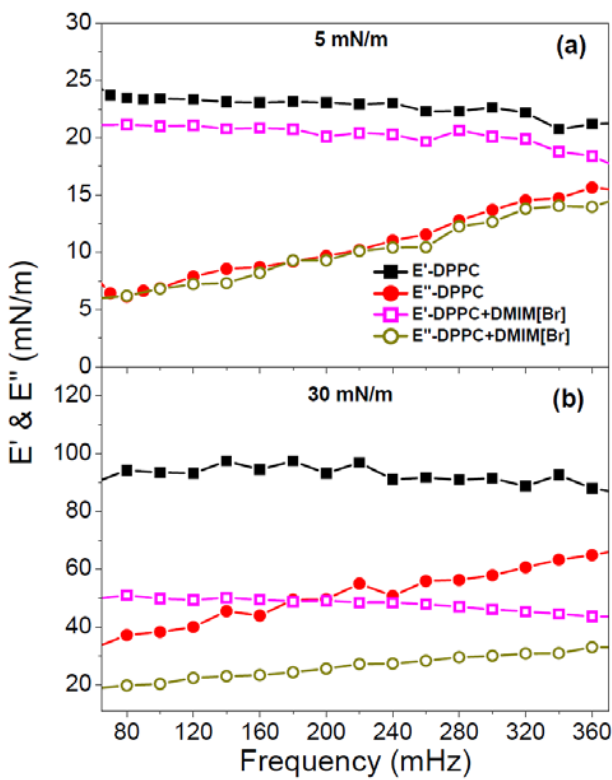

**FIGURE 3:** The storage (*E′*) and loss (*E″*) modulus of DPPC lipid in the absence and presence of 10 wt% DMIM[Br] ILs at two different phases, (a) LE phase at 5 mN/m, and (b) LC phase at 30 mN/m surface pressure.

As DMIM[Br] shows prominent effects on the isotherm and in-plane static elasticity of the DPPC lipid monolayer, in this current investigation, a technique of dilation rheology has been employed on this monolayer in the presence of DMIM[Br] to analyze the dynamic viscoelastic nature of the layers. The measurements are performed to quantify the storage modulus ($E'$) and loss modulus ($E''$) of layer in LE and LC phases keeping the layer at a surface pressure of 5 and 30 mN/m, respectively. As shown in Fig. 3 (a) and (b), in the LE phase, the values of $E'$ is always greater than the $E''$ suggesting the elastic nature of lipid film. While there is slow decrease in $E'$ for both phases, the values of $E''$ rise with frequency. Even the drop of $E'$ is insignificant in addition of 10 wt% of the DMIM[Br] in the lipid layer in LE phase, but it become drastic in the case of LC phase. Here, this drop signifies the enhanced in-plane flexibility of the lipid film. This dynamic result is consistent with the in-plane static elasticity, calculated from the isotherm of the monolayer showing a drop in its value. A similar result for long chain ionic liquid, DMIM[Cl] has been reported for egg-sphingomyelin lipid membrane[47]. It is further interesting to observe that the value of $E''$ also falls in the presence of the IL explaining a lower frictional force felt by a molecule while it tries to diffuse laterally in the plane. Such an observation accommodates well the higher APM observed in the surface area-pressure isotherm shown in Fig. 2. The lower value of $E''$ implies that the motion of the lipid molecules would be faster in the plane of a lipid layer.

All the above isotherm and rheology results collectively suggest a strong interaction between ILs and the monolayers, significantly impacting the intermolecular interactions among DPPC molecules. Moreover, ILs emerge as influential modifiers of the static elastic and dynamic viscoelastic properties of the monolayer, underscoring their significant role in modulating the mechanical behavior of the membrane. Given that ULVs consist of two opposing monolayers forming a bilayer structure, we further investigate the impact of ILs on the DPPC bilayer.

### 3.3 ILs Modulate Phase Behaviour and Induce Conformational Disorder in Membrane

FTIR spectroscopy was used to examine the effects of ILs on the thermotropic phase behavior and conformational changes in DPPC membranes. FTIR analyzes IR-active bands from fundamental vibrations involving $CH_2$ groups (such as C-H stretching and $CH_2$ scissoring modes), which are sensitive to the *trans-gauche* ratio. These bands are ideal for identifying the main phase transition of membranes, as this transition involves the alkyl tails shifting from an all-*trans* state to a disordered state with significant *gauche* defects i.e., conformational disorder. Consequently, FTIR provides information on the lipid molecule's conformation during phase transitions, effectively complementing the DSC which tracks changes in the system's enthalpy during phase transitions.

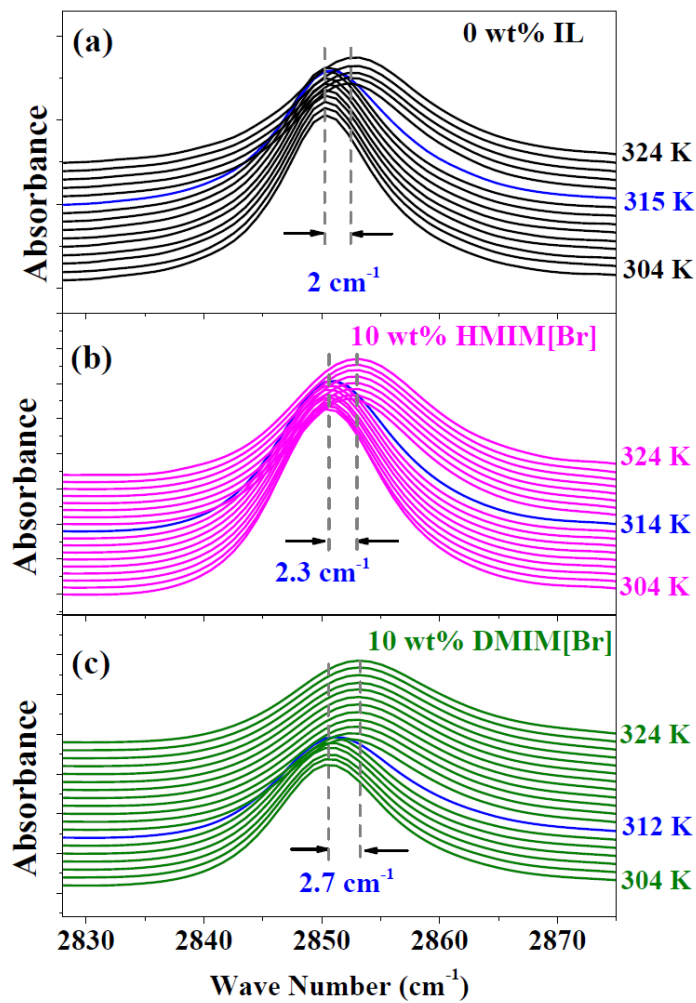

**FIGURE 4:** FTIR absorbance spectra for DPPC with: (a) 0 wt% IL (black), (b) 10 wt% HMIM[Br] (magenta), and (c) 10 wt % DMIM[Br] (green). Dashed lines are the guide for the band centers. Spectra where the sudden blueshift occurs is highlighted by blue color.

Our FTIR analysis focuses on the $CH_2$ symmetric stretching mode ($\upsilon_s(CH_2)$) in the alkyl chains to monitor change in tail region. A linear correlation has been shown between the conformational order of the alkyl chain and methylene stretching vibrations[48]. Stretching frequencies for the *gauche* conformer involve large energy and occur at a relatively higher frequency compared to the *trans* conformer[49-50]. At the main phase transition, an abrupt increase in the frequency (i.e. blue shift) and broadening of the $\upsilon_s(CH_2)$ are observed[51]. Variations in the peak centers of the $\nu_s(CH_2)$ mode serve as an indicator of alterations in the *gauche/trans* ratio in lipid chains[18]. This information is closely linked to the state of conformation order/disorder, influencing the phase transition behavior of membranes. Peak width is associated with the rate of *trans-gauche* isomerization. Figure 4(a-c) shows the temperature-dependent FTIR absorbance spectra of the $\nu_s(CH_2)$ mode in DPPC membranes with varying ILs concentrations. The spectral analysis of this mode spans the wave number range of 2830-2875 cm$^{-1}$. It is evident that as the temperature increases, peak gets broadened and blue shifted.

For quantitative analysis, temperature-dependence of peak centers of $\nu_s(CH_2)$ mode for DPPC membrane at different concentrations of ILs are shown in Figure 5 (a). For neat DPPC membrane, at 305 K, the peak center of the $\nu_s(CH_2)$ mode is found at ~ 2850.7 cm$^{-1}$. This peak center indicates the prevalence of all-*trans* conformations in the alkyl chains of DPPC, suggesting the formation of an ordered gel phase[18-19]. As the temperature rises, a sharp blueshift (~2 cm$^{-1}$) in peak centers is observed at 315 K, marking the $T_m$, where ordered gel phase undergoes a transition into the disordered fluid phase, as observed in the DSC thermogram[7, 23]. This blueshift in peak center signifies the emergence of *gauche* defects in the alkyl chains of DPPC within the membrane. The presence of *gauche* defects in the fluid phase makes the membrane highly disordered compared to a membrane in the gel phase. It is evident that the incorporation of 10 wt % HMIM[Br] in the DPPC ULVs marginally shifts $T_m$ towards lower temperature, 314 K. Furthermore, it slightly blueshifts the $\nu s(CH_2)$ mode in both the gel and fluid phases suggesting HMIM[Br] IL create disorder in the membrane in both gel and fluid phases. Upon further increase of HMIM[Br] concentration to 20 wt %, the blueshift in $\nu_s(CH_2)$ mode becomes more pronounced, especially in the fluid phase but doesn't affect $T_m$ much. In the case of longer alkyl chain length IL, DMIM[Br], effects are more significant. At 10 wt % of the IL, $T_m$ substantially decreases to 311 K. Furthermore, a more pronounced blueshift is observed at peak centers in both the phases due to the presence of DMIM[Br] in DPPC ULVs as compared to the presence of HMIM[Br] at the same concentration. This suggests that a longer alkyl chain of IL creates more disruption in the membrane. Increasing the DMIM[Br] concentration to 20 wt% further enhances the blueshift in both the phases and shifts $T_m$ to an even lower temperature of 309 K. In summary, the $T_m$, exhibits a decrease and blue shift in peak centers of $\nu_s(CH_2)$ increases upon incorporation of both the ILs, with the effect being consistently more pronounced with longer IL DMIM[Br] and higher concentration (20 wt%). These findings align closely with insights from DSC studies[23].

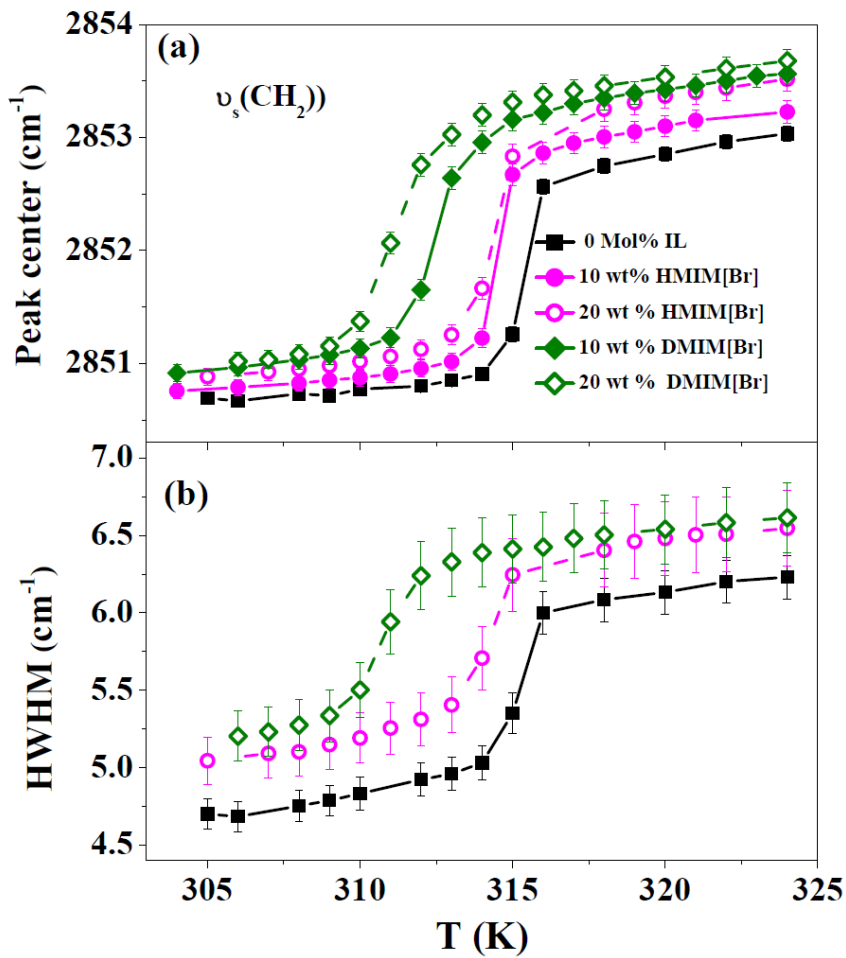

**FIGURE 5:** (a) Temperature-dependent peak centers of $\nu_s(CH_2)$ mode in the alkyl chains of DPPC with 0 wt% (black), 10 wt% (filled symbols), and 20 wt% (open symbols) HMIM[Br] (magenta) and DMIM[Br] (green) ILs. (b) Temperature-dependent HWHM of $\nu_s(CH_2)$ mode in the alkyl chains of DPPC with 0 and 20 wt% ILs.

Temperature-dependent variations in the half width at half maxima (HWHM) of the absorbance peak in the $\nu_s(CH_2)$ mode, influenced by the presence of 20 wt % ILs in DPPC membrane, are shown in Fig. 5 (b). At low temperature, a small HWHM (4.6 cm$^{-1}$) is observed for the pure DPPC membrane in the gel phase. As the temperature increases, the HWHM increases. The origin of the increase in HWHM is the augmentation of the rotational/conformational motion of the hydrocarbon chains. A sharp jump in the HWHM is noted at 315 K, suggesting increased mobility of this CH$_2$ group at the main phase transition. Therefore, in the fluid phase, dynamics of the alkyl chains becomes faster, surpassing those observed in the gel phase. Upon the incorporation of ILs into the DPPC membrane, we observed a consistent increase in HWHM across the entire temperature range. This observation indicates that the presence of ILs enhances the dynamics of the alkyl chains in the DPPC membrane. The increase in chain dynamics contribute to augmented membrane fluidity, evident in both gel and fluid phases.

### 3.4. ILs Accelerates Membrane Dynamics at Nanoscales

To investigate the effects of ILs on the microscopic dynamics of the membrane, QENS experiments were conducted on DPPC membranes with and without HMIM[Br] and DMIM[Br] at 300 K and 345 K, where the DPPC membrane is in the gel and fluid phases, respectively. Measurements were carried out at different concentrations of DMIM[Br]. These experiments elucidate the role of the membrane's physical state, concentration, and alkyl chain length of IL on the microscopic dynamics of lipid membranes. In QENS experiments involving hydrogenous systems like lipid membranes, incoherent scattering from hydrogen atoms dominates the spectra due to hydrogen's high incoherent scattering cross section compared to other atoms (C, N, P, O, etc.). To enhance the signal from the lipid membrane, $D_2O$ was used as the solvent. Figure 6 shows the observed QENS spectra at 310 K for DPPC vesicle solutions and $D_2O$ at $Q = 1.0$ Å$^{-1}$. The QENS spectra from vesicle solution represent the sum of contributions from both the membrane and the solvent, allowing the membrane contribution to be extracted as[18-19]

$$S_{mem}(Q,E) = S_{solution}(Q,E) - \phi S_{solvent}(Q,E) \tag{6}$$

where $S_{mem}(Q, E)$, $S_{solution}(Q, E)$ and $S_{solvent}(Q, E)$ are the scattering functions for DPPC lipid membrane, vesicles solution and solvent, respectively. The factor $\phi$ accounts for the volume fraction of solvent ($D_2O$) in the solution. Representative DPPC spectra are shown in Fig. 6, with the instrument resolution from the vanadium standard in the inset. Significant quasielastic broadening, exceeding the instrument resolution, indicates stochastic lipid motion at both the temperatures. A clear increase in quasielastic broadening is observed when transitioning from the gel phase (310 K) to the fluid phase (345 K).

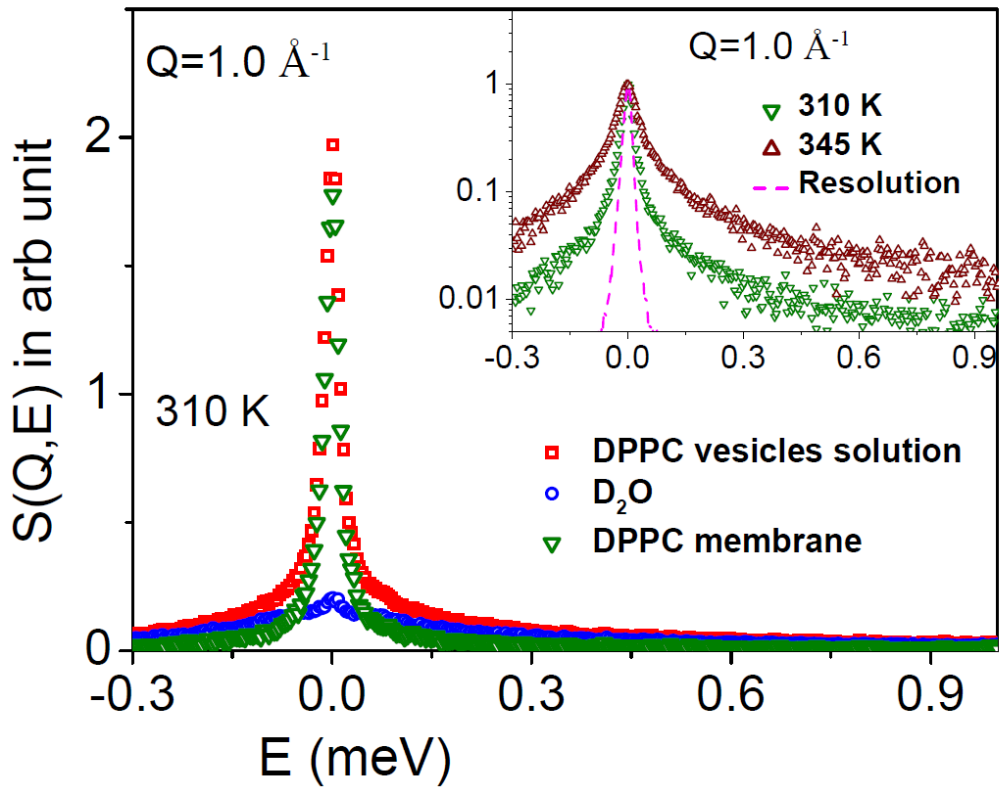

**FIGURE** 6 : Representative QENS spectra for DPPC vesicle solution, $D_2O$, and DPPC membrane ($D_2O$ contribution subtracted) at 310 K and $Q$ = 1.0 Å$^{-1}$. The inset shows QENS spectra for the DPPC membrane in the gel phase (310 K) and fluid phase (345 K) at $Q$ = 1.0 Å$^{-1}$. Instrument resolution, measured from a vanadium standard, is indicated by a dashed line. For quantitative comparison, the spectra are normalized to peak amplitude.

Figure 7(a) and (b) illustrate the QENS spectra for the DPPC membrane at varying concentrations of DMIM[Br], measured at a $Q$ value of 1.2 Å$^{-1}$, for the gel and fluid phases, respectively. To enable the direct comparison, QENS spectra have been normalized by their peak amplitudes. In both phases, the incorporation of ILs significantly enhances quasielastic broadening, indicating increased membrane dynamics. Higher concentrations of ILs further accelerate membrane dynamics, suggesting that ILs function as effective plasticizers.

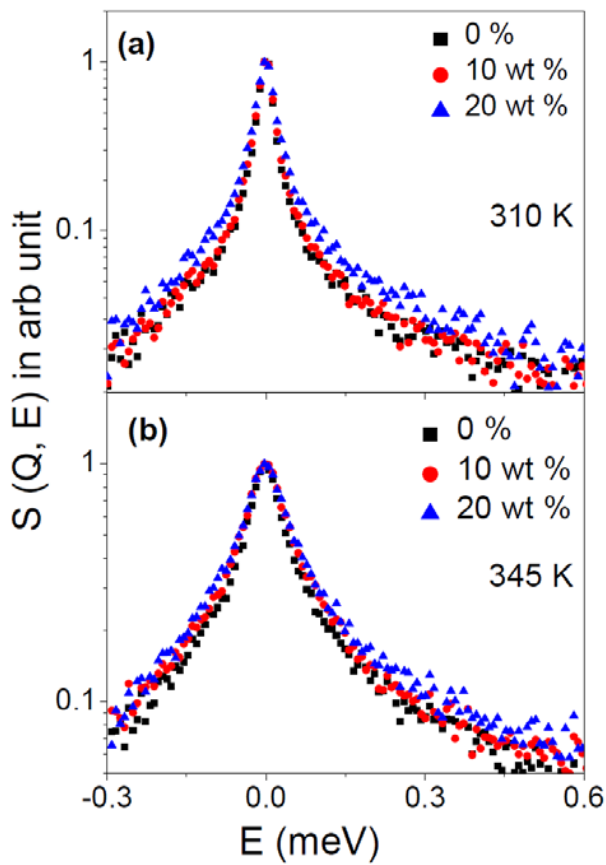

**FIGURE** 7: Typical QENS Spectra at $Q$=1.2 Å$^{-1}$ for the DPPC membrane with varying DMIM[Br] concentrations at (a) 310 K (gel phase) and (b) 345 K (fluid phase). The contribution of the solvent (D$_2$O) has been subtracted, and the resultant spectra are normalized to the peak amplitudes.

QENS probes system dynamics on nanoseconds to picoseconds timescales and Angstroms to nanometers length scales. In this domain, lipid molecules exhibit two main dynamical modes: (i) lateral motion within the leaflet and (ii) localized internal motions of lipid alkyl chains[52]. The mechanism of lateral diffusion remains debated, with models such as continuous diffusion, ballistic motions, localized diffusion, and sub-diffusive motion[29-30, 53-54]. Our study adopts a simple continuous diffusion model, effective for describing motions over distances greater than a lipid's diameter[28, 30]. The associated scattering law for lateral motion follows a Lorentzian distribution. Internal lipid motions, constrained by chemical structure, add an elastic component to Lorentzian component. Thus, the overall scattering law for a lipid membrane is a convolution of the scattering laws for lateral and internal motions, and can be written as[28]

$$S_{mem}(Q,E) = A(Q)L_{lat}(\Gamma_{lat},E) + (1 - A(Q)L_{tot}(\Gamma_{lat} + \Gamma_{int},E) \tag{7}$$

where $A(Q)$ represents the elastic incoherent structure factor (EISF) of internal motion, and $\Gamma_{lat}$ and $\Gamma_{int}$ denote the HWHM of the Lorentzian functions corresponding to lateral and internal motions of the lipid. Equation (7) was convoluted with the instrument resolution function, and the values of $A(Q)$, $\Gamma_{lat}$, and $\Gamma_{int}$ were obtained through least-square fitting of the QENS spectra. QENS data analysis was conducted using the DAVE software[55] developed at the NIST Center for Neutron Research.

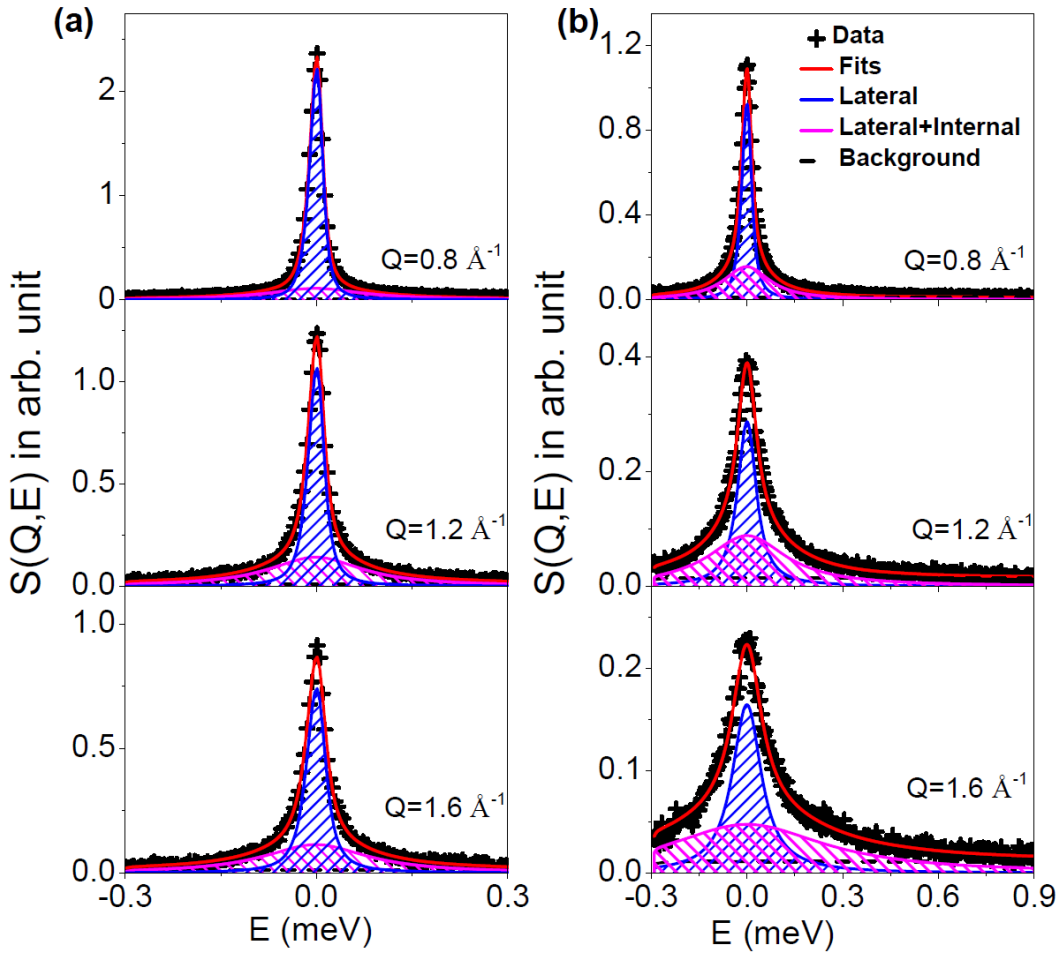

**FIGURE** 8 :Typical fitted QENS spectra for DPPC membrane with 10 wt % DMIM[Br] at (a) 310 K and (b) 345 K with the model scattering function given by Eq. (7). Individual components corresponding lateral and combination of lateral and internal are also shown.

The scattering law described by Equation (7) describes the QENS data well for DPPC membranes with and without ILs at both temperatures. Figure 8(a) and (b) represent typical fitted QENS spectra for the DPPC membrane with 10 wt% DMIM[Br] at different $Q$ values at 310 K and 345 K, respectively, showing the components for lateral motion and the combination of lateral and internal motions. To gain quantitative insights, the $Q$-dependences of the fitting parameters $A(Q)$, $\Gamma_{lat}$, and $\Gamma_{int}$ were further analyzed.

### 3.4.1 Lateral Diffusion

Lateral diffusion of lipids is crucial for the bilayer's response time to external perturbations, impacting cell signaling, membrane trafficking, protein-protein interactions, and the localization of membrane proteins. Changes in lateral motion can affect membrane fluidity, permeability, and transport properties. In liposome-based drug delivery systems, lipid lateral diffusion influences the stability and release kinetics of drugs. The obtained HWHMs associated with the lateral diffusion of lipids from the fitting of QENS data are shown in Figure 9(a) for the DPPC membrane at varying concentrations of DMIM[Br] at 310 K and 345 K. The incorporation of DMIM[Br] significantly enhances $\Gamma_{lat}$ values, indicating faster lateral motion at both temperatures. $\Gamma_{lat}$ is found to increase with the concentration of DMIM[Br], indicating that higher IL concentrations significantly enhance the lateral motion in both ordered and fluid phases. To explore the effect of alkyl chain length, QENS experiments were also conducted on DPPC with a shorter alkyl chain IL, HMIM[Br], at the same concentration as DMIM[Br]. The inset illustrates the variations in HWHMs for DPPC with 10 wt% HMIM[Br] at 310 K and 345 K. The results reveal that, at the same concentration, the shorter alkyl chain IL, HMIM[Br], induces a comparatively smaller acceleration in lateral motion than DMIM[Br]. This observation suggests that the alkyl chain length of ILs plays a crucial role in modulating their impact on membrane dynamics. It is interesting to note that for all the systems and temperatures, $\Gamma_{lat}$ increases linearly with $Q^2$, passing through the origin, indicating that lipid lateral motion follows continuous diffusion as per Fick's law. The lateral diffusion coefficients ($D_{lat}$), derived from the slopes of these linear fits, are presented in Figure9 (b), demonstrating that both ILs enhance lateral diffusion in both the phases. In case of DMIM[Br], a concentration-dependent increase in lateral diffusion coefficient is evident at both the temperatures. For example, at 345 K, $D_{lat}$ increases by 24 % with 10 wt % DMIM[Br] and by 41% with 20 wt% DMIM[Br], highlighting the significant impact of longer-chain IL, DMIM[Br], on enhancing lipid mobility. In comparison, 10 wt% HMIM[Br] increases $D_{lat}$ by only 8%, underscoring the superior efficacy of longer alkyl chain IL, DMIM[Br] in promoting lateral diffusion. This is consistent with the results of ξ-potentials and pressure-area isotherm studies which showed stronger binding and a larger increase in area per lipid for longer alkyl chain IL, DMIM[Br]. QENS results also suggest that IL-membrane interaction varies with the membrane's physical state. For example, at 10 wt% DMIM[Br], $D_{lat}$ increases by 36 % at 310 K (gel phase) and 24% at 345 K (fluid phase). Various theoretical approaches including free volume model[56], Saffman–Delbrück hydrodynamic model[57] have been used to explain the lipid lateral diffusion and shown that this dynamic process is affected by various factors such as the area per lipid, solvent viscosity, and obstruction effects. The addition of ILs disrupts the well-ordered, densely packed lipid molecules in the gel phase, resulting in a large increase in area per lipid that significantly accelerates lateral lipid diffusion, whereas the already disordered fluid phase experiences a less substantial change in area per lipid. This explains dependence of physical state of the membrane on the increase in lateral diffusion of lipid.

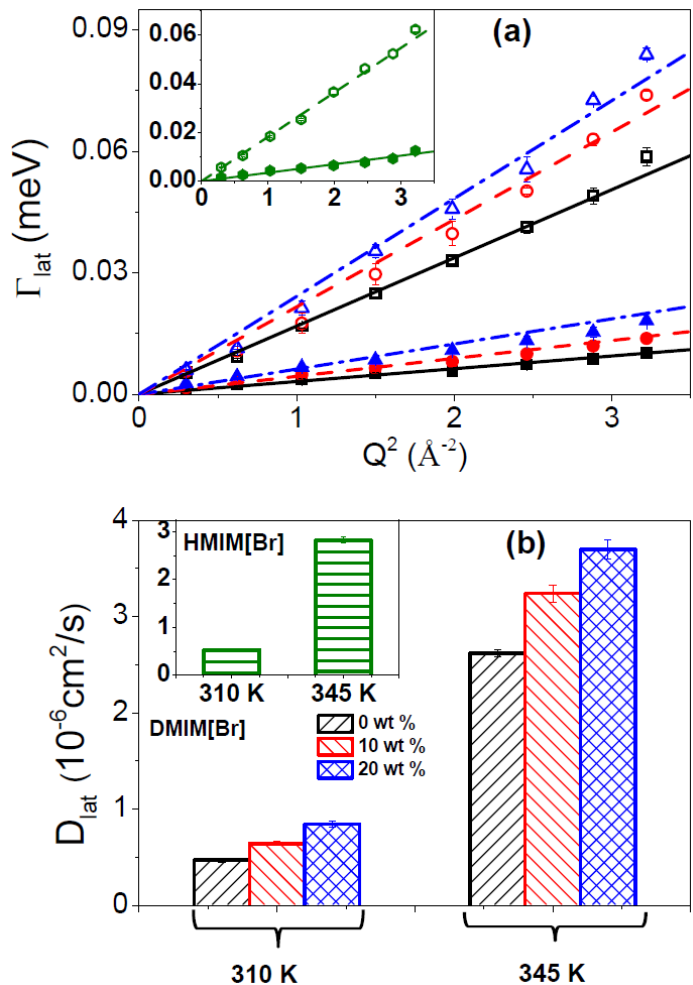

**FIGURE 9:** (a) Variation of the half-width at half-maximum (HWHM) of the Lorentzian representing lateral lipid motion ($\Gamma_{lat}$) with respect to $Q^2$ for DPPC membranes. $\Gamma_{lat}$ is shown for neat DPPC (black squares), DPPC with 10 wt% DMIM[Br] (red circles), and DPPC with 20 wt% DMIM[Br] (blue triangles) at 310 K (filled) and 345 K (open). The inset presents $\Gamma_{lat}$ values for DPPC with 10 wt% HMIM[Br]. The lines represent the Fickian diffusion model discussed in the text. (b) Lateral diffusion coefficient ($D_{lat}$) for DPPC membranes with varying concentrations of DMIM[Br] IL at 310 K and 345 K, illustrating a concentration-dependent increase in lateral motion at both the temperatures. The inset shows the lateral diffusion coefficient for the DPPC membrane with 10 wt% HMIM[Br].

### 3.4.2 Internal Motion

Another type of motion within the relevant timescale of QENS is the internal dynamics of lipid molecules. These motions are crucial for selective permeability and the adjustment of lateral membrane organization, facilitating protein-protein interactions. In liposome drug carriers, internal lipid motion is vital for shaping drug delivery functionalities, such as creating efficient drug-trapping regions and promoting effective drug loading. The flexibility of alkyl chains, driven by internal lipid motion, accommodates diverse-sized drug molecules. Variations of EISF and HWHM corresponding to internal motion for DPPC membranes, with and without ILs, are shown in Fig. 10.

In the ordered phase, lipid molecules exhibit a tightly packed arrangement with alkyl chains predominantly in an all-*trans* conformation. In this configuration, the alkyl chains prefer to undergo uniaxial rotation around their axis within a circular path defined by the radius of gyration ($R$). It is plausible that at a specific temperature, not all hydrogen atoms within the acyl chains exhibit mobility. Consequently, the generalized scattering law describing the internal motion of lipid molecules in the ordered phase can be expressed[58]

$$S_{\text{int}}^{uni}(Q,E) = p_x \delta(E) + \left(1 - p_x\right)\left[ A_0(QR)\delta(E) + \frac{1}{\pi}\sum_{n=1}^{N_s-1} A_n(QR)\frac{\tau_n}{1+E^2\tau_n^2} \right] \tag{8}$$

with

$$A_n(QR) = \frac{1}{N_s}\sum_{i=1}^{N_s} j_0\left( 2QR\sin\frac{\pi i}{N_s} \right)\cos\frac{2\pi ni}{N_s}$$

and

$$\tau_n^{-1} = 2\tau^{-1}sin^2(\frac{n\pi}{N_s}).$$

where, $p_x$ represents the fraction of hydrogen atoms with no observed dynamics within the given timescale, $j_0$ stands for the spherical Bessel function of the zeroth order, $N_s$ denotes the number of sites evenly distributed on a circle with radius $R$, and $\tau$ is the average time spent on a site between two consecutive jumps. For a substantial number of sites ($Ns \geq 6$) and a limited range of $QR$ ($QR \leq \pi$), it has been demonstrated[58] that above scattering law given in Eq. (8) applies to uniaxial rotational diffusion. In such cases, the rotational diffusion constant $D_R$ can be expressed as:

$$D_R = \frac{2}{\tau}\sin^2\left( \frac{\pi}{N_s} \right) \tag{9}$$

EISF for the ordered phase can be expressed as

$$(EISF)_{ordered} = p_x + (1-p_x)\frac{1}{N_s}\sum_{i=1}^{N_s} j_0(2QR\sin\frac{\pi i}{N_s}) \tag{10}$$

The observed EISF for the DPPC membrane in the absence and presence of ILs at 310 K was characterized using a least-squares fitting method. Eq. (10), with fit parameters $p_x$ and $R$, was employed for this purpose. The fitting curves presented in Fig. 10 (a) demonstrate that the fractional uniaxial rotational diffusion model could describe the data for all the membrane systems. Obtained values of fitting parameters, the fraction of mobile hydrogen (1-$p_x$) and the radius of rotation ($R$) are given in Table-1. Percentage of mobile hydrogen and radius of rotation are slightly enhanced due to incorporation of DMIM[Br], suggesting an increase in the flexibility of the alkyl chain. No significant change in internal motion is observed due to incorporation of HMIM[Br].

The variation of $\Gamma_{int}$ with $Q$, extracted from the QENS data, is shown in Fig. 10 (b) and appears relatively flat which is a typical signature for the rotational diffusion. The rotational diffusion constant, $D_R$, can be determined through the least squares fitting of $\Gamma_{int}$ using Eqs. (8) and (9). In this fitting process, the value of $R$, obtained from the $Q$-dependence of EISF, is employed as a fixed parameter. For the DPPC membrane, the obtained $D_R$ is $70 \pm 2$ µeV, and this value shows negligible change upon the incorporation of both ILs. These results suggest that unlike to lateral motion, internal motion of DPPC is less sensitive due to presence of IL in the ordered phase of membrane.

**Table: 1** Parameters corresponding to fractional uniaxial rotational diffusion of lipid for DPPC membrane in the absence and presence of ILs in the ordered phase.

| Lipid membrane | Fraction of mobile hydrogen's $(1-p_x)$ (%) | Radius, $R$ (Å) | $D_R$ ( µeV) |
|---|---|---|---|
| DPPC | $57 \pm 2$ | $1.9 \pm 0.1$ | $70 \pm 2$ |
| DPPC+10 wt % DMIM[Br] | $57 \pm 2$ | $1.8 \pm 0.1$ | $71 \pm 2$ |
| DPPC+20 wt % DMIM[Br] | $63 \pm 3$ | $2.0 \pm 0.1$ | $73 \pm 2$ |
| DPPC+10 wt % HMIM[Br] | $58 \pm 3$ | $1.9 \pm 0.1$ | $70 \pm 2$ |

In the fluid phase, lipid molecules exhibit loose packing, and their alkyl chains display substantial *gauche* defects, resulting in a larger area per molecule. In this state, internal motions such as reorientations, conformational changes, bending, and torsional motions occur. These motions can be modeled by assuming each hydrogen atom undergoes localized translational diffusion (LTD) within spherical domains. This model aligns with the data in Fig. 10(b), where $\Gamma_{int}$ shows a finite non-zero value as $Q{\rightarrow}0$ and increases monotonically at higher $Q$. We have adopted a linear distribution for the radii and diffusivities, similar to other long-chain molecular systems[18-19]. The resulting scattering law can be written as[28, 59]

$$S_{int}^{LTD}(Q,E) = \frac{1}{N_c} \sum_{i=1}^{N_c} \left[ \left[ \frac{3 j_1(QR_i)}{QR_i} \right]^2 \delta(E) + \frac{1}{\pi} \sum_{\{l,n\} \neq \{0,0\}} (2l+1) A_n^l(QR_i) \frac{(x_n^l) D_i / R_i^2}{\left[ (x_n^l) D_i / R_i^2 \right]^2 + E^2} \right] \quad (11)$$

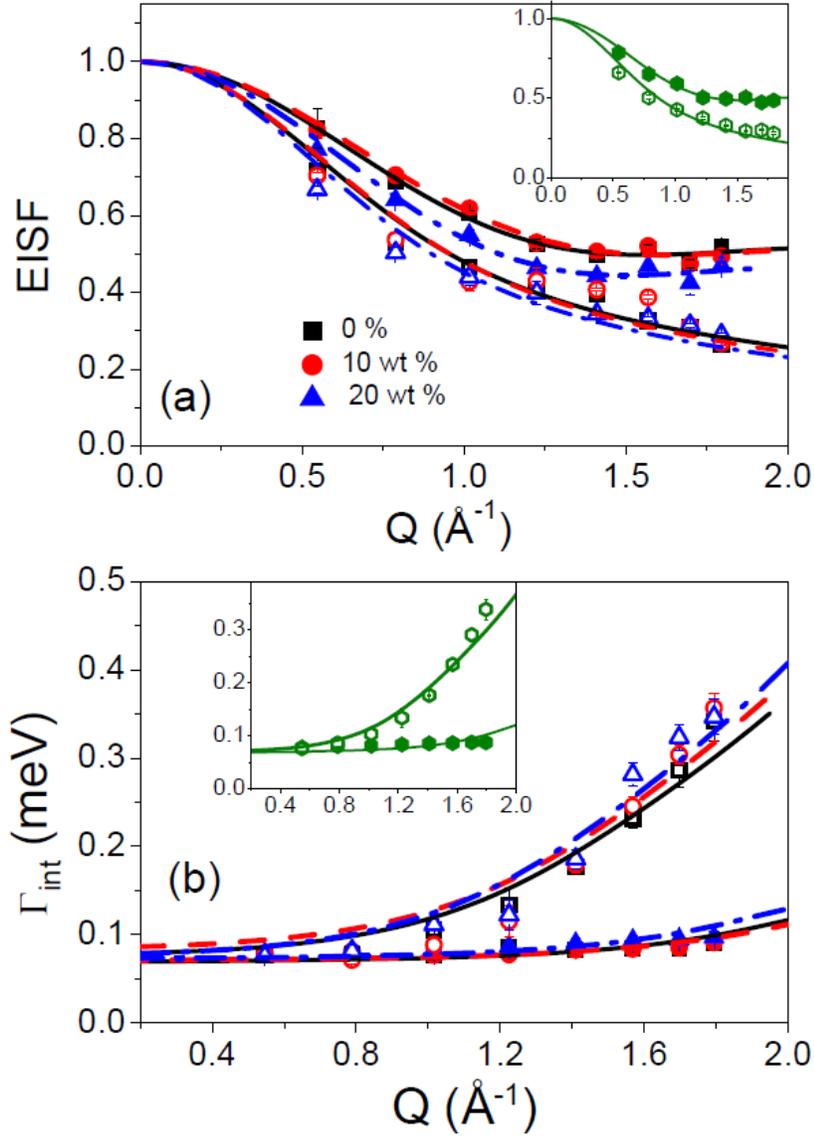

**FIGURE** 10 : Variation of the (a-c) EISF and (d) Lorentzian's HWHM ($\Gamma_{int}$) with $Q$ for the DPPC membrane at different concentrations of DMIM[Br] IL at 310 K (filled symbols) and 345 K (open symbols). Insets show EISF or HWHM data for the DPPC membrane with 10 wt% HMIM[Br]. The lines represent fits using the fractional uniaxial rotational model at 310 K and the localized translational diffusion model at 345 K, as described in the text.

where $N_c$ represents the total number of CH$_2$ units in the alkyl chain and $A_n^1(QR_i)$ is the quasielastic structure factor. This factor can be calculated for different $n$ and $l$ using the values of $x_n{}^l$ listed in ref[59]. $R_i$ and $D_i$ denote the radius of the sphere and the diffusivity associated with the $i^{th}$ site of the alkyl chain and can be expressed as:

$$R_i = \frac{i-1}{N_c-1}\left[R_{\max}-R_{\min}\right]+R_{\min}$$
$$D_i = \frac{i-1}{N_c-1}\left[D_{\max}-D_{\min}\right]+D_{\min} \tag{12}$$

$N_c$ equals to 16 for DPPC lipid. The EISF for the fluid phase can be expressed as:

$$(EISF)_{fluid} = \frac{1}{16}\sum_{i=1}^{16}\left[\frac{3j_i(QR_i)}{QR_i}\right]^2 \tag{13}$$

The EISF for the DPPC membrane, both with and without DMIM[Br] and HMIM[Br], at 345 K was analyzed using a least-squares fitting method. Equations (12) and (13) were utilized for fitting, with $R_{\min}$ and $R_{\max}$ as fit parameters. The fitting curves, shown in Fig. 10 (a), demonstrate the LTD model's efficacy in describing the data for both membrane systems. The $R_{\min}$ values are unrealistically small, suggesting the near immobility of hydrogen atoms in the CH$_2$ units near the head group. The corresponding $R_{\max}$ values are detailed in Table 2. The timescale of internal motion was determined using a subsequent least-squares fitting method with Eqs. (11) and (12), using $D_{\min}$ and $D_{\max}$ as parameters. These fits are represented by lines in Fig. 10 (b). During the fitting of $\Gamma_{int}$, the radii values were fixed based on the $Q$ dependence of EISF. The model accurately describes the behavior of $\Gamma_{int}(Q)$, as shown in Fig. 10 (b), with the corresponding parameters also provided in Table 2.

**Table: 2** Parameters corresponding to localised translational diffusion of lipid for DPPC membrane in absence and presence of ILs in the fluid phase.

| Fluid | System | $R_{\max}$ (Å) | $D_{\min}$ ($\times 10^{-6}$ cm$^2$/s) | $D_{\max}$ ($\times 10^{-6}$ cm$^2$/s) |
|---|---|---|---|---|
| | DPPC | $3.9 \pm 0.2$ | $7.6 \pm 0.2$ | $32.1 \pm 0.8$ |
| 345 K | DPPC+10 wt % DMIM[Br] | $3.9 \pm 0.3$ | $9.1 \pm 0.2$ | $33.1 \pm 0.9$ |

| | DPPC+20 wt % DMIM[Br] | $4.1 \pm 0.3$ | $10.6 \pm 0.3$ | $33.9 \pm 0.9$ |
|---|---|---|---|---|
| | DPPC+10 wt % HMIM[Br] | $4.1 \pm 0.4$ | $9.1 \pm 0.3$ | $32.0 \pm 0.8$ |

In the fluid phase, the introduction of ILs slightly increases the size of spherical domains and their associated diffusivities, indicating a minimal impact on the internal motion of lipids. This aligns with NMR studies on zwitterionic POPC membranes, which show a decrease in lipid chain order parameter and elasticity with IL addition, regardless of IL chain length[22].

### 3.5 Insights from MD Simulation

To gain deeper microscopic insights, MD simulations were performed on DPPC membranes with varying concentrations of DMIM[Br]. These simulations were conducted at seven different temperatures, spanning the main phase transition. In accordance with the experimental conditions, the ILs were initially mixed and equilibrated with the lipid bilayer in its fluid phase for approximately 100 ns. During this period, a significant fraction of ILs successfully integrated into the lipid membrane. Subsequently, the system temperature was systematically decreased through simulated annealing. Equilibrated snapshots of DPPC with DMIM[Br] at two different temperatures, 307 K (below $T_m$) and 326 K (above $T_m$), representing the gel and fluid phases, respectively, are shown in Figure 11 (a) & (b). In the gel phase, the acyl chains exhibit a high degree of order and are densely packed, whereas in the fluid phase, the acyl chains become highly disordered, resulting in an increased area per lipid (APL). Interestingly, even at 307 K, where the membrane predominantly exists in the gel phase, most ILs remained embedded within the membrane. This behavior stands in stark contrast to the scenario where ILs were introduced directly into the gel phase, as the tightly packed and highly ordered lipid structure in this phase impeded significant IL incorporation[7]. Equilibrated APL has been calculated using the following expression

$$APL = \frac{2(L_x \times L_y)}{N_{lipid}} \qquad (14)$$

where $L_x$, $L_y$ are the equilibrated simulation box length in the $x$ and $y$ directions respectively and $N_{lipid}$ is the total number of lipids in the simulation. The factor of 2 accounts for the presence of lipids in both leaflets of the bilayer. The calculated APL values for DPPC membranes at various temperatures are presented in Figure 11 (c). At 303 K, the APL is relatively low, as expected, since the DPPC membrane exists in the tightly packed gel phase. With increasing temperature, the APL exhibits a gradual rise, followed by a sharp increase around 315 K, marking the characteristic signature of the main phase transition. Notably, our MD simulations accurately capture the main phase transition temperature of pure DPPC lipids. Upon the addition of 4 wt% DMIM[Br], a notable increase in APL is observed in both the gel and fluid phases, aligning well with small-angle neutron scattering (SANS) studies[23], which reported bilayer thinning due to IL incorporation. This is also consistent with the results of pressure-area isotherm which suggest addition of IL increases the APL. Furthermore, the main phase transition temperature shifts to a lower value, indicating a disruption in the lipid packing. As the IL concentration increases, the phase transition temperature continues to shift downward, accompanied by a further increase in APL in both phases. These findings are consistent with the results of FTIR, which suggest that increasing IL concentration induces more *gauche* defects in the acyl chains, leading to greater lipid disorder and a subsequent rise in APL. In order to substantiate this, *gauche/trans* ratio is calculated from MD simulation trajectories, by classifying the dihedrals along the alkyl chain of the lipids. Dihedrals falling in the range $160°–200°$ were categorized as *trans* conformation, wherein dihedrals in range $40° -80°$ and $280°-320°$ were classified as *gauche* conformers. Based on this classification, the obtained *gauche/trans* ratio along the dihedrals in the alkyl chain is shown in Fig. 11(d). Quite evidently, the addition of 10 mol% DMIM[Br] increases the number of *gauche* conformers, particularly near the middle of the chain. This feature can be ascribed to the increased volume availability due to the penetration of ILs into the lipid membrane. These results are consistent with FTIR results which suggest incorporation of IL blue shifts in peak center of $\nu_s(CH_2)$ suggesting more *gauche/trans* ratio.

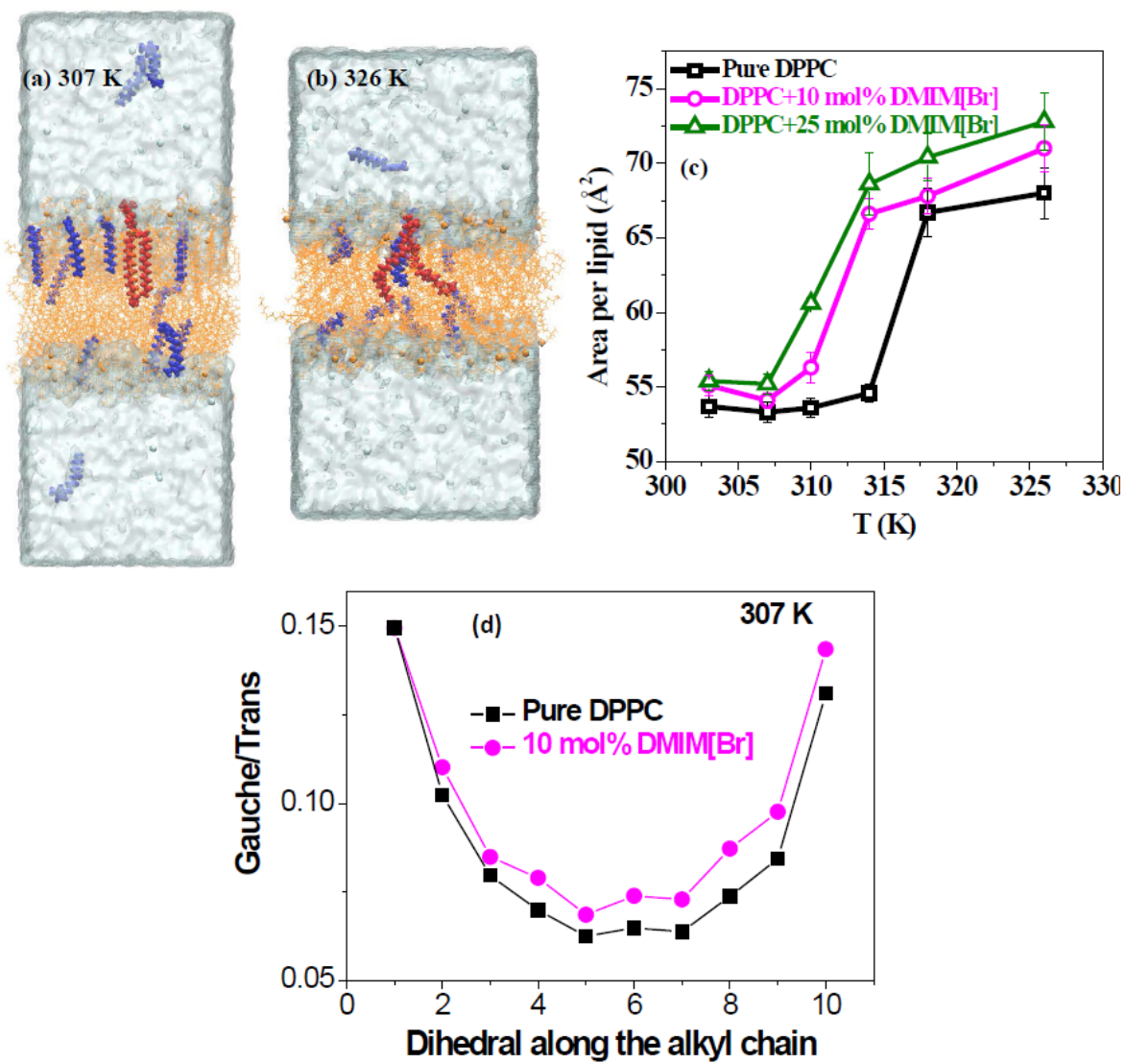

**FIGURE** 11: Snapshots of the DPPC+DMIM[Br] simulation at (a) 307 K (gel phase) and (b) 326 K (fluid). Lipids are shown in yellow, DMIM$^+$ in blue. (c) Variation of area per lipid (APL) of DPPC with temperature at varying concentrations of DMIM[Br]. (d) *Gauche* to *trans* ratio in alkyl chains of DPPC in absence and presence of DMIM[Br].

The free volume model suggests that lateral lipid diffusion is strongly influenced by the APL[56, 60]. According to this model, lateral diffusion is governed by the availability of free volume exceeding a critical size adjacent to a diffusing lipid molecule. An increase in APL generally facilitates faster lateral diffusion by providing more space for lipid movement within the membrane. Our QENS results clearly demonstrate that the incorporation of ILs enhances lateral lipid diffusion, with the degree of enhancement increasing as IL concentration rises. This behavior can be directly correlated with the concentration-dependent increase in APL, as revealed by our MD simulations. The observed relationship underscores the crucial role of ILs in modulating membrane dynamics at the molecular level. Maintaining optimal membrane fluidity is essential for cellular function, ensuring the proper movement of lipids and other components while preserving membrane integrity and preventing the uncontrolled leakage of substances.

Even subtle changes in membrane fluidity can have profound effects on cellular stability. Our findings suggest that the presence of ILs significantly accelerates lipid lateral diffusion, which could potentially compromise membrane integrity by increasing permeability. Such disruptions may lead to enhanced ion transport or lipid scrambling, posing a serious threat to cellular homeostasis.

## 4. CONCLUSIONS

This study provides a comprehensive understanding of the interactions between imidazolium-based ionic liquids (ILs) and lipid membranes, with significant implications for both their toxicological properties and potential pharmaceutical applications. Using dipalmitoylphosphatidylcholine (DPPC) membranes as a model system, we demonstrate that ILs profoundly influence membrane viscoelasticity, dynamics, and structural organization.

Our study reveals that ILs induce substantial structural disorder in lipid membranes, evident from increased area per lipid molecules and disrupted packing. This disorder is further pronounced with ILs having longer alkyl chains, which exhibit stronger interactions with the membrane. The impact of ILs extends to the membrane's phase behavior, as observed through shifts in the main phase transition to lower temperatures and the introduction of *gauche* defects, reflecting increased flexibility and reduced structural integrity.

Additionally, ILs significantly enhance the lateral diffusion ($D_{lat}$) of lipids within the membrane, an effect that is strongly influenced by the physical state of the membrane, the concentration of ILs, and the length of their alkyl chains. The most pronounced enhancement in lipid dynamics is observed in ordered membrane phases at higher concentrations of longer-chain ILs. These findings suggest that ILs disrupt the ordered state of the membrane more effectively than the fluid state, leading to enhanced lateral mobility of lipid molecules.

Molecular dynamics (MD) simulations corroborate the experimental results, revealing that ILs induce disruption in lipid organization, increase the area per lipid, and thereby significantly enhance lateral diffusion. This disruption in membrane organization and dynamics is likely to increase membrane fluidity and permeability, providing a mechanistic basis for the higher toxicity associated with ILs.

The study underscores the critical role of IL structure and concentration in determining their effects on membrane properties. The insights gained here highlighted the delicate balance between IL-induced beneficial properties, such as enhanced permeability for drug delivery, and their potential cytotoxicity due to excessive disruption of membrane structure and function.

These findings pave the way for designing ILs with tailored properties for specific pharmaceutical applications, emphasizing the need for a detailed understanding of IL-membrane interactions to optimize their safety and efficacy. The other broader implication of this study is towards the efficient drug delivery system. Liposomes emerge as highly promising carriers for drug delivery, and the stability, as well as release kinetics of these liposomes, crucially hinge on the phase behaviour and membrane fluidity. Our findings signify that the incorporation of ILs plays a pivotal role in modulating both the phase behaviour and fluidity of the membrane. These factors dictate the balance of electrostatic and hydrophobic interactions, and are therefore instrumental in enhancing the efficiency of cargo transport. This insight underscores the potential of IL-modified liposomes as effective platforms for advanced drug delivery systems.

## 5. ACKNOWLEDGMENTS

The authors sincerely thank Dr. V. G. Sakai and Dr. H. Bhatt for their help with QENS & FTIR measurements, respectively.